\documentclass[titlepage,12pt]{article}
\usepackage{amssymb}
\usepackage{amsfonts}

\begin{document}

\title{Scattering and bound states of spinless particles in a mixed
vector-scalar smooth step potential}
\date{}
\author{M.G. Garcia, A.S. de Castro\thanks{%
E-mail address: castro@pq.cnpq.br} \\
%EndAName
\\
UNESP - Campus de Guaratinguet\'{a}\\
Departamento de F\'{\i}sica e Qu\'{\i}mica\\
12516-410 Guaratinguet\'{a} SP - Brazil}
\maketitle

\begin{abstract}
Scattering and bound states for a spinless particle in the background of a
kink-like smooth step potential, added with a scalar uniform background, are
considered with a general mixing of vector and scalar Lorentz structures.
The problem is mapped into the Schr\"{o}dinger-like equation with an
effective Rosen-Morse potential. It is shown that the scalar uniform
background present subtle and trick effects for the scattering states and
reveals itself a high-handed element for formation of bound states. In that
process, it is shown that the problem of solving a differential equation for
the eigenenergies is transmuted into the simpler and more efficient problem
of solving an irrational algebraic equation.
\end{abstract}

\section{Introduction}

The one-dimensional step potential is of certain interest to model the
transition between two structures. In solid state physics, for example, a
step-like potential which changes continuously over an interval whose
dimensions are of the order of the interatomic distances can be used to
model the average potential which holds the conduction electrons in metals.
In the presence of strong potentials, though, the Schr\"{o}dinger equation
must be replaced by their relativistic counterparts. The scattering of
spin-1/2 particles by a square step potential, considered as a time
component of a vector potential, is well-known and crystalized in textbooks
\cite{gre}. In that scenario it appears the celebrated Klein's paradox \cite{kle}. The analysis of the same problem with the Klein-Gordon
(KG) equation was not neglected \cite{gro}, \cite{wi}. The background of the
kink configuration of the $\phi ^{4}$ model ($\mathrm{tanh}\,\gamma x$) \cite%
{raj} is of interest in quantum field theory where topological classical
backgrounds are responsible for inducing a fractional fermion number on the
vacuum. Models of these kinds, known as kink models are obtained in quantum
field theory as the continuum limit of linear polymer models \cite{gol}. In
a recent paper the complete set of bound states of fermions in the presence
of this sort of kink-like smooth step potential has been addressed by
considering a pseudoscalar coupling in the Dirac equation \cite{asc}. A
peculiar feature of the kink-like potential is the absence of bounded
solutions in a nonrelativistic theory because it gives rise to an ubiquitous
repulsive potential. Of course this problem neatly reveals that our
nonrelativistic preconceptions are mistaken.

It is well known from the quarkonium phenomenology that the best fit for
meson spectroscopy is found for a convenient mixture of vector and scalar
potentials put by hand in the equations (see, e.g. \cite{luc}). The same can
be said about the treatment of the nuclear phenomena describing the
influence of the nuclear medium on the nucleons \cite{ser}. It happens that
when the vector and scalar potentials fulfill the conditions for spin and
pseudospin symmetries, i.e. they have the same magnitude, the energy
spectrum does not depend on the spinorial structure, being identical to the
spectrum of a spinless particle \cite{diracKG4}.

In the present work the scattering a spinless particle in the background of
a kink-like smooth step potential, added with a scalar uniform background,
is considered with a general mixing of vector and scalar Lorentz structures.
Although the scalar potential finds many of their applications in nuclear
and particle physics, it could also simulate an effective mass term in solid
state physics and so it could be useful for modelling transitions between
structures such a Josephson junctions \cite{bra}. It is often useful,
because of simplicity, to approximate the behavior of relativistic fermions
by spinless particles obeying the KG equation. It turns out that some
results almost do not depend on the spin structure of the particle, e.g. the
onset of scaling in some structure functions in the case of relativistic
quark models used for studying quark-hadron duality \cite{diracKG}, the
photoelectron spectra in the strong field laser-induced ionization and
recollision process \cite{diracKG2}, the electric polarizability of the
ground state of a particle bound in a strong Coulomb field \cite{lee}, and
the differential scattering cross section for forward scattering \cite%
{diracKG3}. Nevertheless, our purpose is to investigate the basic nature of
the phenomena without entering into the details involving specific
applications. In other words, the aim of this paper is to search new
solutions of a fundamental equation in physics which can be of help to see
more clearly what is going on into the details of a more specialized and
complex circumstance. In passing, it is shown that a serious problem with
the square step potential, overlooked in the literature, does not manifest
for the smooth step potential. Our problem is mapped into an exactly
solvable Sturm-Liouville problem of a Schr\"{o}dinger-like equation with an
effective Rosen-Morse potential which been applied in discussing polyatomic
molecular vibrational states \cite{ros}. The scalar uniform background makes
its influence not only for the scattering states but reveals itself a
high-handed element for formation of bound states. In that process, the
problem of solving a differential equation for the eigenenergies is
transmuted into the simpler and more efficient problem of solving an
irrational algebraic equation. With this methodology the whole relativistic
spectrum is found, if the particle is massless or not. Nevertheless, bounded
solutions do exist only under strict conditions.

\section{The KG equation with vector and scalar potentials}

The (1+1)-dimensional KG equation for a free particle of rest mass $m$
cor\-res\-pon\-ding to the relativistic energy-momentum relation $%
E^{2}=c^{2}p^{2}+m^{2}c^{4}$, where the energy $E$ and the momentum $p$ are
substituted by operators, $i\hbar \,\partial /\partial t$ and $-i\hbar
\,\partial /\partial x$ respectively, acting on the wave function $\Phi
(x,t) $. Here, $c$ is the speed of light and $\hbar $ is the Planck constant
($\hbar =h/(2\pi )$). In the presence of external potentials the
energy-momentum relation becomes

\begin{equation}
\left( E-V_{t}\right) ^{2}=c^{2}\left( p-\frac{V_{sp}}{c}\right) ^{2}+\left(
mc^{2}+V_{s}\right) ^{2}  \label{1}
\end{equation}%
where the subscripts for the potentials denote their properties under a
Lorentz transformation: $t$ and $sp$ for the time and space components of a
vector potential, and $s$ for the scalar potential. A continuity equation
for the KG equation

\begin{equation}
\frac{\partial \rho }{\partial t}+\frac{\partial J}{\partial x}=0
\label{con}
\end{equation}

\noindent is satisfied with $\rho $ and $J$ defined as

\begin{eqnarray}
\rho &=&\frac{i\hbar }{2mc^{2}}\left( \Phi ^{\ast }\frac{\partial \Phi }{%
\partial t}-\frac{\partial \Phi ^{\ast }}{\partial t}\Phi \right) -\frac{%
V_{t}}{mc^{2}}\left\vert \Phi \right\vert ^{2}  \nonumber \\
&&  \label{rho&J1} \\
J &=&\frac{\hbar }{2im}\left( \Phi ^{\ast }\frac{\partial \Phi }{\partial x}-%
\frac{\partial \Phi ^{\ast }}{\partial x}\Phi \right) -\frac{V_{sp}}{mc}%
\left\vert \Phi \right\vert ^{2}  \nonumber
\end{eqnarray}

\noindent Note that the KG equation is covariant under the
charge-conjugation operation, meaning that the KG equation remains invariant
under the simultaneous transformations $\Phi \rightarrow \pm \Phi ^{\ast }$,
$E\rightarrow -E$, $p\rightarrow -p$, $V_{sp}\rightarrow -V_{sp}$ and $%
V_{t}\rightarrow -V_{t}$. In other words, if $\Phi $ is a solution for a
particle (antiparticle) with energy $E$ and momentum $p$ for the potentials $%
V_{t}$, $V_{sp}$ and $V_{s}$, then $\pm \Phi ^{\ast }$ is a solution for a
antiparticle (particle) with energy $-E$ and momentum $-p$ for the
potentials $-V_{t}$, $-V_{sp}$ and $V_{s}$. Note also that $\rho \rightarrow
-\rho $ and $J\rightarrow -J$. These last results are of particular
importance to interpret $\rho $ and $J$ as charge density and charge current
density, respectively, and to recognize that the vector potential couples
with the charge of the particle/antiparticle whereas the scalar potential
couples with the mass, as one could suspect from the appearance of $V_{s}$
in (\ref{1}) and from the absence of $V_{s}$ in (\ref{rho&J1}). Furthermore,
the change $E\rightarrow -E$ and related change $i\hbar \,\partial /\partial
t\rightarrow -$ $i\hbar \,\partial /\partial t$ permit us to conclude that
if the particle travels forward in time then the antiparticle travels
backward in time.

For time-independent potentials the KG equation admits solutions in the form

\begin{equation}
\Phi (x,t)=\varphi (x)\,e^{\frac{i}{\hbar c}\Lambda \left( x\right) }\,e^{-%
\frac{i}{\hbar }Et}  \label{PHI}
\end{equation}

\noindent where $\,\varphi $ satisfies the time-independent KG equation

\begin{equation}
-\frac{\hbar ^{2}}{2m}\,\varphi ^{\prime \prime }+V_{\mathtt{eff}}\,\varphi
=E_{\mathtt{eff}}\,\varphi  \label{VEFF}
\end{equation}

\noindent with

\begin{equation}
E_{\mathtt{eff}}=\frac{E^{2}-\left( mc^{2}\right) ^{2}}{2mc^{2}},\quad V_{%
\mathtt{eff}}=\frac{V_{s}^{2}-V_{t}^{2}}{2mc^{2}}+V_{s}+\frac{E}{mc^{2}}%
\,V_{t}  \label{eff}
\end{equation}

\noindent and

\begin{equation}
\Lambda \left( x\right) =\int^{x}d\eta \,V_{sp}\left( \eta \right)
\label{Lambda}
\end{equation}

\noindent The density and flux corresponding to (\ref{PHI}) are then

\begin{equation}
\rho =\frac{E-V_{t}}{mc^{2}}\left\vert \varphi \right\vert ^{2},\quad J=%
\frac{\hbar }{2im}\left( \varphi ^{\ast }\frac{\partial \varphi }{\partial x}%
-\frac{\partial \varphi ^{\ast }}{\partial x}\varphi \right)  \label{RHO}
\end{equation}

\noindent Since $\rho $ and $J$ are independent of time, $\varphi $ is said
to describe a stationary state. Notice that the density becomes negative in
regions of space where $V_{t}>E$, so that the KG wave function must be
normalized as

\begin{equation}
\int_{-\infty }^{+\infty }dx\,\frac{E-V_{t}(x)}{mc^{2}}\left\vert \varphi
(x)\right\vert ^{2}=\pm 1  \label{normarho}
\end{equation}

\noindent where the $\pm $ sign must be used for

\begin{equation}
E\gtrless \frac{\int_{-\infty }^{+\infty }dx\,V_{t}(x)\left\vert \varphi
(x)\right\vert ^{2}}{\int_{-\infty }^{+\infty }dx\,\left\vert \varphi
(x)\right\vert ^{2}}  \label{ineqE}
\end{equation}

\noindent Meanwhile, in the nonrelativistic approximation (potential
energies small compared to $mc^{2}$ and $E\simeq mc^{2}$) Eq. (\ref{VEFF})
becomes

\begin{equation}
\left( -\frac{\hbar ^{2}}{2m}\frac{d^{2}}{dx^{2}}+V_{t}+V_{s}\right) \varphi
=\left( E-mc^{2}\right) \varphi  \label{1e}
\end{equation}

\noindent so that $\varphi $ obeys the Schr\"{o}dinger equation with binding
energy equal to $E-mc^{2}$ without distinguishing the contributions of
vector and scalar potentials. Furthermore, the density and current (and the
normalization condition with the plus sign too) reduce precisely to the
corresponding values of the nonrelativistic theory.

It is remarkable that the KG equation with a scalar potential, or a vector
potential contaminated with some scalar coupling, is not invariant under $%
V\rightarrow V+\mathrm{const.}$, this is so because the vector potential
couples to the charge of the particle, whereas the scalar potential couples
to the mass of the particle. Therefore, if there is any scalar coupling the
absolute values of the energy will have physical significance and the
freedom to choose a zero-energy will be lost. As we will see explicitly in
this work, a constant added to the scalar potential is undoubtedly
physically relevant. As a matter of fact, it can play a crucial role to
ensure the existence of bound-state solutions even though the bound states
are not present in the nonrelativistic limit of the theory. It is well known
that a binding potential in the nonrelativistic approach is not binding in
the relativistic approach when it is considered as a Lorentz vector. It is
not immediately obvious that relativistic binding potentials may only result
in scattering states in the nonrelativistic approach. The secret lies in the
fact that vector and scalar potentials couple differently in the KG equation
whereas there is no such distinction among them in the Schr\"{o}dinger
equation. This observation permit us to conclude that even a
\textquotedblleft repulsive\textquotedblright\ potential might be a bona
fide binding potential.

\section{The mixed vector-scalar kink-like potential and the effective
Rosen-Morse potential}

Now let us focus our attention on scalar and vector potentials in the form
of smooth steps:

\begin{eqnarray}
V_{t} &=&g_{t}V,\quad V_{s}=g_{s}\left( \mathcal{V}+V\right)  \nonumber \\
&&  \label{4} \\
\mathcal{V} &=&\mathrm{const},\quad V=\frac{V_{0}}{2}\left( 1+\tanh \,\frac{x%
}{2L}\right)  \nonumber
\end{eqnarray}

\noindent where the dimensionless coupling constants, $g_{t}$ and $g_{s}$,
are real numbers constrained by $g_{t}+g_{s}=1$. The positive parameter $L$
is related to the range of the interaction which makes $V$ to change
noticeably in the interval $-2L<x<2L$, and as $L\rightarrow 0$ the potential
approximates the square step potential. $V_{0}>0$ is the height of the
potential $V$ at $x=+\infty $. The uniform background makes the height of
the scalar potential at $x=-\infty $ to be $g_{s}\mathcal{V}$. \ The reason
for including a scalar uniform background will be clear later -- it makes
possible the existence of bound-state solutions.

Before proceeding, it is useful to note that the effective potential
corresponding to (\ref{4}) \ is recognized as the exactly solvable
Rosen-Morse potential \cite{ros}-\cite{nie}

\begin{equation}
V_{\mathtt{eff}}=-V_{1}\,\mathrm{sech}^{2}\frac{x}{2L}+V_{2}\,\mathrm{tanh}\,%
\frac{x}{2L}+V_{3}  \label{6}
\end{equation}%
\noindent where the following abbreviations have been used

\[
V_{1}=\left( 2g_{s}-1\right) \frac{V_{0}^{2}}{8mc^{2}}
\]

\begin{equation}
V_{2}=2V_{1}+\frac{V_{0}}{2mc^{2}}\,\left( g_{s}Mc^{2}+g_{t}E\right)
\label{potpar}
\end{equation}

\[
V_{3}=V_{2}+E_{\mathtt{eff}}-\frac{E^{2}-\left( Mc^{2}\right) ^{2}}{2mc^{2}}
\]

\noindent and

\begin{equation}
M=m+g_{s}\frac{\mathcal{V}}{c^{2}}  \label{mef}
\end{equation}

\noindent The Rosen-Morse potential approaches $V_{3}\pm V_{2}$ as $%
x\rightarrow \pm \infty $, and has an extremum when $|V_{2}|<2|V_{1}|$ at

\begin{equation}
x_{m}=L\ln \left( \frac{2V_{1}-V_{2}}{2V_{1}+V_{2}}\right)  \label{xm}
\end{equation}

\noindent given by

\begin{equation}
V_{\mathtt{eff}}\left( x_{m}\right) =V_{3}-V_{1}\left[ 1+\left( \frac{V_{2}}{%
2V_{1}}\right) ^{2}\right]  \label{vmin}
\end{equation}

\noindent The second derivative of $V_{\mathtt{eff}}$ at $x_{m}$ is given by

\begin{equation}
V_{\mathtt{eff}}^{\prime \prime }\left( x_{m}\right) =\frac{1}{\left(
2V_{1}\right) ^{3}}\left[ \frac{\left( 2V_{1}\right) ^{2}-V_{2}^{2}}{2L}%
\right] ^{2}  \label{der2}
\end{equation}

\noindent in such a way that the extremum is a local minimum (maximum) only
if $V_{1}>0 $ ($V_{1}<0$). Note that the extremum, if it exists at all, is a
minimum (maximum) only if $g_{s}>1/2$ ($g_{s}<1/2$). In particular, the
symmetric Rosen-Morse potential is that one with $V_{2}=0$ which can be
obtained with

\begin{equation}
g_{s}^{2}\,\mathcal{V}=-\left[ g_{s}\,mc^{2}+g_{t}\,E+\left( 2g_{s}-1\right)
\frac{V_{0}}{2}\right]  \label{vezinho}
\end{equation}

As a matter of fact, potential-well structures can be achieved when $%
|V_{2}|<2|V_{1}|$ $\ $with $V_{1}>0$ and \ a pressing need for bound-state
solutions \ implies that $E_{\mathtt{eff}}$ defined in (\ref{eff}) must
satisfy $V_{\mathtt{eff}}\left( x_{m}\right) <E_{\mathtt{eff}}<V_{\mathtt{eff%
}}(\pm \infty )$. From the condition $E_{\mathtt{eff}}<V_{\mathtt{eff}}(\pm
\infty )$ one concludes that

\begin{equation}
\left\vert E\right\vert <|M|c^{2},\quad |E-g_{t}V_{0}|<|Mc^{2}+g_{s}V_{0}|
\label{ineq4}
\end{equation}%
Writing $V_{\mathtt{eff}}\left( x_{m}\right) $ as

\begin{equation}
V_{\mathtt{eff}}\left( x_{m}\right) =V_{2}-V_{1}\left[ 1+\left( \frac{V_{2}}{%
2V_{1}}\right) ^{2}\right] +E_{\mathtt{eff}}-\frac{E^{2}-\left(
Mc^{2}\right) ^{2}}{2mc^{2}}  \label{ineq5}
\end{equation}

\noindent one can see that the condition $E_{\mathtt{eff}}>V_{\mathtt{eff}%
}\left( x_{m}\right) $ turns into

\begin{equation}
E\neq -\frac{g_{t}}{g_{s}}Mc^{2}  \label{ineq6}
\end{equation}%
According to this last result, there is no crossing between energy levels
with $E>-g_{t}Mc^{2}/g_{s}$ (to be associated with the particle levels) and
those ones with $E<-g_{t}Mc^{2}/g_{s}$ (to be associated with the particle
levels). This fact implies that there is no channel for spontaneous
particle-antiparticle creation, so that the single-particle interpretation
of the KG equation is preserved. In particular, null energies are not
permissible for bound-state solutions in the case of a pure scalar coupling (%
$g_{s}=1$), when the energy levels are disposed symmetrically about $E=0$.
Furthermore, the coefficient $V_{2}$ in (\ref{6}) can also be expressed as

\begin{equation}
V_{2}=-2V_{1}+\frac{V_{0}}{2mc^{2}}\,\left[ g_{s}Mc^{2}+g_{t}E+\left(
2g_{s}-1\right) V_{0}\right]  \label{new V2}
\end{equation}%
From this and from the definition of $V_{2}$ in (\ref{potpar}), it follows
that

\begin{equation}
g_{s}Mc^{2}+g_{t}E<0  \label{ineq1}
\end{equation}

\noindent and

\begin{equation}
g_{s}Mc^{2}+g_{t}E+\left( 2g_{s}-1\right) V_{0}>0  \label{ineq2}
\end{equation}

\noindent Combining these two inequalities one concludes that an additional
sine qua non condition for the existence of bounded solutions is that $%
Mc^{2} $ must be into the limits

\begin{equation}
-\left( \frac{2g_{s}-1}{g_{s}}V_{0}+\frac{g_{t}}{g_{s}}E\right) <Mc^{2}<-%
\frac{g_{t}}{g_{s}}E  \label{ineq3}
\end{equation}

To acknowledge that the effective potential for the mixing given by (\ref{4}%
) is a Rosen-Morse potential can help you to see more clearly how a
kink-like smooth step potential might furnish bound-state solutions. After
all, we shall not use the knowledge about the exact analytical solution for
the Rosen-Morse potential.

\subsection{\protect The asymptotic solutions}

As $|x|\gg L$ the effective potential is practically constant (the main
transition region occurs in $|x|<2L$) and the solutions for the KG equation
can be approximate by those ones for a free particle. Furthermore, the
asymptotic behaviour will show itself suitable to impose the appropriate
boundary conditions to the complete solution to the problem.

The vector and scalar potentials approaches to zero and to $g_{s}\mathcal{V}$
as $x\rightarrow -\infty $. Hence the solution for the KG equation can be
written as

\begin{equation}
\varphi \left( -\infty \right) =A_{+}e^{+ikx}+A_{-}e^{-ikx}  \label{sol z<0}
\end{equation}

\noindent where

\begin{equation}
\hbar ck=\sqrt{E^{2}-\left( Mc^{2}\right) ^{2}}  \label{k}
\end{equation}%
For $\left\vert E\right\vert >|M|c^{2}$, the solution expressed by (\ref{sol
z<0}) describes plane waves propagating on both directions of the $X$-axis
with group velocity $v_{g}=\left( dE/dk\right) /\hbar $ equal to the
classical velocity. If we consider that particles are incident on the
potential $\left( E>|M|c^{2}\right) $, $A_{+}\exp (+ikx)$ will describe
particles coming to the potential region from $-\infty $ ($v_{g}=+c^{2}\hbar
k/E>0$), whereas $A_{-}\exp (-ikx)$ will describe reflected particles ($%
v_{g}=-c^{2}\hbar k/E<0$). The flux corresponding to $\varphi $ given by (%
\ref{sol z<0}), is expressed as

\begin{equation}
J(-\infty )=J_{\mathtt{inc}}-J_{\mathtt{ref}}  \label{j1}
\end{equation}

\noindent where

\begin{equation}
J_{\mathtt{inc}}=\frac{\hbar k}{m}\left\vert A_{+}\right\vert ^{2},\quad J_{%
\mathtt{ref}}=\frac{\hbar k}{m}\left\vert A_{-}\right\vert ^{2}  \label{jota}
\end{equation}

\noindent Note that the relation $J=\rho v_{g}$ maintains for the incident
and reflected waves, since

\begin{equation}
\rho _{\pm }(-\infty )=\frac{E}{mc^{2}}\left\vert A_{\pm }\right\vert ^{2}
\label{j2}
\end{equation}

\noindent On the other hand, the vector and scalar potentials approaches to $%
V_{0}$ and to $g_{s}\left( \mathcal{V}+V_{0}\right) $ as $x\rightarrow
+\infty $. In this asymptotic region one should have $v_{g}\geq 0$ in such a
way that the solution in this region of space describes an evanescent wave
or a progressive wave running away from the potential potential region. The
general solution has the form

\begin{equation}
\varphi \left( +\infty \right) =B_{+}e^{+i\kappa x}+B_{-}e^{-i\kappa x}
\label{sol z0}
\end{equation}

\noindent where

\begin{equation}
\hbar c\kappa =\sqrt{\left( E-g_{t}V_{0}\right) ^{2}-\left(
Mc^{2}+g_{s}V_{0}\right) ^{2}}  \label{kapa}
\end{equation}%
Due to the twofold possibility of signs for the energy of a stationary
state, the solution involving $B_{-}$ can not be ruled out a priori. As a
matter of fact, this term may describe a progressive wave with negative
charge density and phase velocity $v_{ph}=|E|/\left( \hbar \kappa \right) >0$%
. It is true that if $\kappa \in \mathbb{R}$ the solution describing a plane
wave propagating in the positive direction of the $X$-axis with group
velocity $v_{g}=\pm c^{2}\hbar \kappa /\left( E-g_{t}V_{0}\right) >0$ is
possible only if $E\gtrless g_{t}V_{0}$ with $B_{\mp }=0$. In this case the
density and the flux corresponding to $\varphi $ given by (\ref{sol z0}) are
expressed as

\begin{equation}
\rho (+\infty )=\frac{E-g_{t}V_{0}}{mc^{2}}\left\vert B_{\pm }\right\vert
^{2},\quad J(+\infty )=J_{\mathtt{trans}}=\pm \frac{\hbar \kappa }{m}%
\left\vert B_{\pm }\right\vert ^{2}  \label{c11}
\end{equation}%
If $\kappa $ is imaginary one can write $\kappa =\pm iQ$ with $Q\in \mathbb{R%
}$, and (\ref{sol z0}) with $B_{\mp }=0$ describes an evanescent wave ($%
v_{g}=0)$. The condition $B_{\mp }=0$ is necessary for furnishing a finite
charge density as $x\rightarrow \infty $. In this case

\begin{equation}
\rho (+\infty )=\frac{E-g_{t}V_{0}}{mc^{2}}\left\vert B_{\pm }\right\vert
^{2}e^{-2Qx},\quad J(+\infty )=0  \label{c100}
\end{equation}

\noindent When $\kappa \in\mathbb{R}$, a bizarre circumstance occurs as long
as $E<g_{t}V_{0}$ since both $\rho (+\infty )$ and $J(+\infty )$ are
negative quantities. The maintenance of the relation $J=\rho v_{g}$, though,
is a license to interpret the solution $B_{-}e^{-i\kappa x}$ as describing
the propagation, in the positive direction of the $X$-axis, of particles
with  charges of opposite sign to the incident particles. This
interpretation is consistent if the particles moving in this asymptotic
region have energy $-E$ and are under the influence of a potential $%
-g_{t}V_{0}$. It means that, in fact, the progressive wave describes the
propagation of antiparticles in the positive direction of the $X$-axis. If $%
\kappa $ is imaginary, though, the solution with $E>g_{t}V_{0}$ ($%
E<g_{t}V_{0}$) describes an evanescent wave associated with particles
(antiparticles).

Defining

\begin{equation}
\varepsilon =\frac{g_{t}}{1-g_{t}}|M|c^{2},\quad V_{c}=\left\{
\begin{array}{c}
\frac{E+Mc^{2}}{2g_{t}-1}, \\
\\
\infty ,%
\end{array}%
\begin{array}{c}
\textrm{for\quad }g_{t}>\frac{1}{2} \\
\\
\textrm{for\quad}g_{t}\leq \frac{1}{2}%
\end{array}%
\right.  \label{epsilon}
\end{equation}%
and the following cases

\begin{itemize}
\item Case I. $M>0$

\item Case II. $M<0$ with $\ g_{t}\leq 1/2$

\item Case III. $M<0$ with $\ 1/2<g_{t}<1$ and $E>\varepsilon $

\item Case IV. $M<0$ with $\ 1/2<g_{t}<1$ and $E<\varepsilon $
\end{itemize}

\noindent one can readily envisage that the segregation between $\kappa $
real and $\kappa $ imaginary allows us to identify three distinct class of
scattering solutions for particles depending on $V_{0}$:

\begin{itemize}
\item Class A. $\kappa $ is real with $V_{0}<E-Mc^{2}$ for the cases I, II,
III, and $V_{0}<V_{c}$ for the case IV.

\item Class B. $\kappa $ is imaginary with $E-Mc^{2}<V_{0}<V_{c}$ for the
cases I, II, III, and $V_{c}<V_{0}<E-Mc^{2}$ for the case IV.

\item Class C. $\kappa $ is real with $V_{0}>V_{c}$ for the cases I, III,
and $V_{0}>E-Mc^{2}$ for the case IV.
\end{itemize}

\bigskip

Now we focus attention on the calculation of the reflection ($R$) and
transmission ($T$) coefficients. The reflection (transmission) coefficient
is defined as the ratio of the reflected (transmitted) flux to the incident
flux. Since $\partial \rho /\partial t=0$ for stationary states, one has
that $J$ is independent of $x$. This fact implies that

\begin{equation}
R=\frac{|A_{-}|^{2}}{|A_{+}|^{2}}  \label{R1}
\end{equation}%
\begin{equation}
T=\left\{
\begin{array}{c}
\pm \frac{\kappa }{k}\frac{|B_{\pm }|^{2}}{|A_{+}|^{2}}, \\
\\
0,%
\end{array}%
\begin{array}{c}
\textrm{for }\kappa \in\mathbb{R},\;B_{\mp }=0 \\
\\
\textrm{for }\kappa =\pm iQ%
\end{array}%
\right.  \label{T1}
\end{equation}

\noindent For all the cases one should have \ $R+T=1$, as expected for a
conserved quantity. This fact is easily verified for $\kappa $ imaginary.
For $\kappa $ real one has to wait for the complete solution of the problem
whose asymptotic behaviour allows one to calculate the amplitudes of all
waves relative to amplitude of the incident wave. Is is instructive to note
that the case with $B_{+}=0$ presents $R>1$, the alluded Klein's paradox, implying that more particles are reflected from the potential
region than those incoming. \ Note that for $\mathcal{V}>-mc^{2}/g_{s}$ $%
(M>0)$ the threshold for pair production is equal to $2mc^{2}$ for $g_{t}=1$
and greater than $2mc^{2}$ for $1/2<g_{t}<1$. For $\mathcal{V}<-mc^{2}/g_{s}$
$(M<0,g_{t}<1)$, though, one has that the threshold is equal to $2|M|c^{2}$,
and that the threshold tends to zero as $\mathcal{V}$ tends to $%
-mc^{2}/g_{s} $. In this last circumstance, pair production occurs for every
$V_{0}$, however small.

It is worthwhile to note that the asymptotic solutions with $k=\pm iq$,
where $q\in\mathbb{R}$, and $\kappa =\pm iQ$, might describe the possible
existence of bound states realized beforehand in the previous section, since
$J(\pm \infty )=0$. In this case, one has to impose that $A_{\pm }=0$ for $%
k=\pm iq$.

\subsection{The complete solutions}

Armed with the knowledge about asymptotic solutions and with the definitions
of the reflection and transmission coefficients we proceed for searching
solutions on the entire region of space.

Changing the independent variable $x$ in (\ref{VEFF}) to

\begin{equation}
y=\frac{1}{2}\left( 1-\tanh \frac{x}{2L}\right)   \label{xtoy}
\end{equation}%
the KG equation is transformed into the differential equation for $\varphi
(y)$:

\begin{equation}
y\left( 1-y\right) \varphi ^{\prime \prime }+\left( 1-2y\right) \varphi
^{\prime }+\Theta \varphi =0  \label{eq1}
\end{equation}

\noindent where

\begin{equation}
\Theta =\left( \frac{L}{\hbar c}\right) ^{2}\frac{\left[ E-g_{t}V_{0}\left(
1-y\right) \right] ^{2}-\left\{ mc^{2}+g_{s}\left[ \mathcal{V}+V_{0}\left(
1-y\right) \right] \right\} ^{2}}{y\left( 1-y\right) }  \label{eq1a}
\end{equation}%
Introducing a new function $\psi (y)$ through the relation%
\begin{equation}
\varphi (y)=y^{\nu }\left( 1-y\right) ^{\mu }\psi (y)  \label{rel}
\end{equation}

\noindent and defining

\begin{eqnarray}
a &=&\mu +\nu +\frac{1-\omega }{2},\quad b=\mu +\nu +\frac{1+\omega }{2}%
,\quad C=2\nu +1  \nonumber \\
&&  \label{def} \\
\mu ^{2} &=&-\left( kL\right) ^{2},\quad \nu ^{2}=-\left( \kappa L\right)
^{2},\quad \omega ^{2}=1+\left( 2g_{s}-1\right) \left( \frac{2LV_{0}}{\hbar c%
}\right) ^{2}  \nonumber
\end{eqnarray}%
the equation (\ref{eq1}) becomes the hypergeometric differential equation
\cite{abr}

\begin{equation}
y\left( 1-y\right) \psi ^{\prime \prime }+\left[ C-\left( a+b+1\right) y%
\right] \psi ^{\prime }-ab\psi =0  \label{hyper}
\end{equation}

\noindent whose general solution can be written in terms of the Gauss
hypergeometric series

\begin{equation}
_{2}F_{1}\left( a,b,C,y\right) =\frac{\Gamma \left( C\right) }{\Gamma \left(
a\right) \Gamma \left( b\right) }\sum\limits_{n=0}^{\infty }\frac{\Gamma
\left( a+n\right) \Gamma \left( b+n\right) }{\Gamma \left( C+n\right) }\frac{%
y^{n}}{n!}  \label{Gauss hyper}
\end{equation}

\noindent in the form \cite{abr}

\begin{equation}
\psi =A\;_{2}F_{1}\left( a,b,C,y\right) +By^{-2\nu }\;_{2}F_{1}\left(
a+1-C,b+1-C,2-C,y\right)   \label{gen1}
\end{equation}

\noindent in such a way that

\[
\varphi =A\,y^{\nu }\left( 1-y\right) ^{\mu }\;_{2}F_{1}\left(
a,b,C,y\right)
\]

\begin{equation}
+B\,y^{-\nu }\left( 1-y\right) ^{\mu }\;_{2}F_{1}\left(
a+1-C,b+1-C,2-C,y\right)   \label{phi1}
\end{equation}

\noindent with the constants $A$ and $B$ to be fitted by the asymptotic
behaviour analyzed in the previous discussion.

As $x\rightarrow +\infty $ (that is, as $y\rightarrow 0$), one has that $%
y\simeq \exp \left( -x/L\right) $ and (\ref{phi1}), because $_{2}F_{1}\left(
a,b,C,0\right) =1$, reduces to

\begin{equation}
\varphi \left( +\infty \right) \simeq Ae^{-\nu x/L}+Be^{\nu x/L}
\label{phi9}
\end{equation}

\noindent so the asymptotic behaviour requires that $B=0$, and $A=B_{\pm }$
for $\nu =\mp i\kappa L$. The asymptotic behaviour as $x\rightarrow -\infty $
($y\rightarrow 1$) can be found by using the relation for passing over from $%
y$ to $1-y$:

\[
_{2}F_{1}\left( a,b,C,y\right) =\gamma _{-}\;_{2}F_{1}\left(
a,b,a+b-C+1,1-y\right)
\]

\begin{equation}
+\,\,\gamma _{+}\;_{2}F_{1}\left( C-a,C-b,C-a-b+1,1-y\right) \left(
1-y\right) ^{C-a-b}  \label{cona}
\end{equation}%
where $\gamma _{+}$ and $\gamma _{-}$ are expressed in terms of the gamma
function as

\begin{equation}
\gamma _{-}=\frac{\Gamma \left( C\right) \Gamma \left( C-a-b\right) }{\Gamma
\left( C-a\right) \Gamma \left( C-b\right) },\quad \gamma _{+}=\frac{\Gamma
\left( C\right) \Gamma \left( a+b-C\right) }{\Gamma \left( a\right) \Gamma
\left( b\right) }\quad  \label{g1}
\end{equation}

\noindent which can also be written as

\begin{eqnarray}
\gamma _{-} &=&\frac{\Gamma \left( 2\nu +1\right) \Gamma \left( -2\mu
\right) }{\Gamma \left( \frac{1+\omega }{2}+\nu -\mu \right) \Gamma \left(
\frac{1-\omega }{2}+\nu -\mu \right) }  \nonumber \\
&&  \label{g20} \\
\gamma _{+} &=&\frac{\Gamma \left( 2\nu +1\right) \Gamma \left( 2\mu \right)
}{\Gamma \left( \frac{1+\omega }{2}+\nu +\mu \right) \Gamma \left( \frac{%
1-\omega }{2}+\nu +\mu \right) }\quad  \nonumber
\end{eqnarray}%
Now, as $x\rightarrow -\infty $, $1-y\simeq \exp \left( +x/L\right) $. This
time, (\ref{phi1}) tends to

\begin{equation}
\varphi \left( -\infty \right) \simeq A\gamma _{+}e^{-\mu x/L}+A\gamma
_{-}e^{+\mu x/L}  \label{phi10}
\end{equation}

\noindent so that $A\gamma _{\pm }=A_{\pm }$ for $\mu =-ikL$, in accordance
with the previous analysis for very large negative values of $x$.

Therefore, the asymptotic behaviour of the general solution dictates that $%
B=0$ and establishes conditions on $\mu $ and $\nu $, but not on $\omega $.
The reflection (\ref{R1}) and transmission (\ref{T1}) coefficients can now
be expressed as

\begin{equation}
R=\frac{|\gamma _{-}|^{2}}{|\gamma _{+}|^{2}}=\frac{|\Gamma \left( \frac{%
1+\omega }{2}+\nu +\mu \right) \Gamma \left( \frac{1-\omega }{2}+\nu +\mu
\right) |^{2}}{|\Gamma \left( \frac{1+\omega }{2}+\nu -\mu \right) \Gamma
\left( \frac{1-\omega }{2}+\nu -\mu \right) |^{2}}\frac{|\Gamma \left( -2\mu
\right) |^{2}}{|\Gamma \left( 2\mu \right) |^{2}}  \label{r2}
\end{equation}

\begin{equation}
T=\left\{
\begin{array}{c}
\frac{\nu }{\mu }\frac{1}{|\gamma _{+}|^{2}}=\frac{\nu }{\mu }\frac{|\Gamma
\left( \frac{1+\omega }{2}+\nu +\mu \right) \Gamma \left( \frac{1-\omega }{2}%
+\nu +\mu \right) |^{2}}{|\Gamma \left( 2\nu +1\right) \Gamma \left( 2\mu
\right) |^{2}}, \\
\\
0,%
\end{array}%
\begin{array}{c}
\textrm{for\quad }\nu =\mp i\kappa L \\
\\
\textrm{for\quad }\nu \in\mathbb{R}%
\end{array}%
\right.  \label{t2}
\end{equation}%
In the numerical evaluation of $R$ and $T$ one has not only to distinguish
the sign of the imaginary part of $\nu $ but also if or not $\omega $ is
real. The following \ identities involving the gamma function \cite{abr}

\begin{equation}
\Gamma \left( z^{\ast }\right) =\Gamma ^{\ast }\left( z\right) ,\quad \Gamma
\left( z\right) \Gamma \left( 1-z\right) =\frac{\pi }{\sin \left( \pi
z\right) }  \label{gama1}
\end{equation}

\noindent are sufficient enough to show that

\begin{equation}
|\Gamma \left( u+iv\right) \Gamma \left( 1-u+iv\right) |^{2}=\frac{2\pi ^{2}%
}{\cosh \left( 2\pi v\right) -\cos \left( 2\pi u\right) }  \label{gama2}
\end{equation}

\noindent where $u$ and $v$ are the real and imaginary parts of $z$.
Furthermore, the following identities will be useful \cite{abr}

\begin{equation}
|\Gamma \left( iv\right) |^{2}=\frac{\pi }{v\sinh \left( \pi v\right) }%
,\quad |\Gamma \left( 1+iv\right) |^{2}=\frac{\pi v}{\sinh \left( \pi
v\right) }  \label{gama3}
\end{equation}

\noindent Hence, one can find for $\mu =-ikL$ and $\nu =\mp i\kappa L$:

\begin{equation}
R=\left\{
\begin{array}{c}
\frac{\cosh \left[ 2\pi \left( k\mp \kappa \right) L\right] -\cos \left[ \pi
\left( 1+\omega \right) \right] }{\cosh \left[ 2\pi \left( k\pm \kappa
\right) L\right] -\cos \left[ \pi \left( 1+\omega \right) \right] }\lessgtr
1, \\
\\
\frac{\cosh \left[ \pi \left( k\mp \kappa +N\right) L\right] \cosh \left[
\pi \left( k\mp \kappa -N\right) L\right] }{\cosh \left[ \pi \left( k\pm
\kappa -N\right) L\right] \cosh \left[ \pi \left( k\pm \kappa +N\right) L%
\right] }\lessgtr 1,%
\end{array}%
\begin{array}{c}
\textrm{for\quad }\omega \in\mathbb{R} \\
\\
\textrm{for\quad }\omega =2iNL,\;N\in\mathbb{R}%
\end{array}%
\right.  \label{r3}
\end{equation}

\begin{equation}
T=\left\{
\begin{array}{c}
\pm \frac{2\sinh \left( 2\pi kL\right) \sinh \left( 2\pi \kappa L\right) }{%
\cosh \left[ 2\pi \left( k\pm \kappa \right) L\right] -\cos \left[ \pi
\left( 1+\omega \right) \right] }\gtrless 0, \\
\\
\pm \frac{\sinh \left( 2\pi kL\right) \sinh \left( 2\pi \kappa L\right) }{%
\cosh \left[ \pi \left( k\pm \kappa -N\right) L\right] \cosh \left[ \pi
\left( k\pm \kappa +N\right) L\right] }\gtrless 0,%
\end{array}%
\begin{array}{c}
\textrm{for\quad }\omega \in\mathbb{R} \\
\\
\textrm{for\quad }\omega =2iNL,\;N\in\mathbb{R}%
\end{array}%
\right.  \label{t3}
\end{equation}

\noindent whereas for $\mu =-ikL$ and $\nu \in\mathbb{R}$ ($\kappa $ pure
imaginary) one has $T=0$ and

\begin{equation}
R=\frac{|\Gamma \left( \frac{1+\omega }{2}+\nu -ikL\right) \Gamma \left(
\frac{1-\omega }{2}+\nu -ikL\right) |^{2}}{|\Gamma \left( \frac{1+\omega }{2}%
+\nu +ikL\right) \Gamma \left( \frac{1-\omega }{2}+\nu +ikL\right) |^{2}}%
=1,\;\forall \,\omega  \label{r4}
\end{equation}

At any circumstance, from the hyperbolic trigonometric identities involving $%
\cosh \left( z_{1}+z_{2}\right) $, one can easily show that $R+T=1$ and that
as $\omega \rightarrow 1$ one finds:

\begin{equation}
R=\left\{
\begin{array}{c}
\left( \frac{k\mp \kappa }{k\pm \kappa }\right) ^{2}, \\
\\
1,%
\end{array}%
\begin{array}{c}
\\
\textrm{for\quad }\mu =-ikL,\;\nu =\mp i\kappa L \\
\\
\textrm{for\quad }\mu =-ikL,\;\nu \in \mathbb{R}%
\end{array}%
\right.   \label{15}
\end{equation}

\begin{equation}
T=\left\{
\begin{array}{c}
\pm \frac{4k\kappa }{\left( k\pm \kappa \right) ^{2}}, \\
\\
0,%
\end{array}%
\begin{array}{c}
\textrm{for\quad }\mu =-ikL,\;\nu =\mp i\kappa L \\
\\
\textrm{for\quad }\mu =-ikL,\;\nu \in\mathbb{R}%
\end{array}%
\right.  \label{16}
\end{equation}

\noindent Note that (\ref{15}) and (\ref{16}) reduce to the results for the
square step potential \cite{gro}, \cite{wi}, as they should be since $\omega
\rightarrow 1$ as $L\rightarrow 0$. Now, $R$ and $T$ blow up for $\kappa =k$
(Class C), i.e. when (\ref{vezinho}) is satisfied. Of course, this crisis
never mentioned in the literature does not mean that the KG theory fails. It
only means that the calculations lose their validity for discontinuous
potentials.

\subsection{Bound states}

As we have said, the solution expressed by (\ref{phi1}) with $\mu $ and $\nu
$ as real quantities ($k$ and $\kappa $ as imaginary numbers), viz.

\begin{eqnarray}
\mu &=&\frac{L}{\hbar c}\sqrt{\left( Mc^{2}\right) ^{2}-E^{2}}  \nonumber \\
&&  \label{munu} \\
\nu &=&\frac{L}{\hbar c}\sqrt{\left( Mc^{2}+g_{s}V_{0}\right) ^{2}-\left(
E-g_{t}V_{0}\right) ^{2}}  \nonumber
\end{eqnarray}

\noindent might describe the possible existence of bound states by imposing
that $B=0$ and $\gamma _{+}=0$. In view of \ (\ref{g1}) one has to locate
the singular points of $\Gamma \left( a\right) \Gamma \left( b\right) $. It
happens that $\Gamma \left( z\right) $ has simple poles only on the real
axis at $z=-n$ ($n=0,1,2,\ldots $), and invoking the expression of $a$ and $%
b $ in terms of $\mu $, $\nu $ and $\omega $ given by (\ref{def}) one
concludes that $\omega $ has got to be a real number. Evidently $LV_{0}$
must be chosen such that

\begin{equation}
LV_{0}<\left\{
\begin{array}{c}
\infty , \\
\\
\frac{\hbar c}{2\sqrt{1-2g_{s}}},%
\end{array}%
\begin{array}{c}
\textrm{for\quad }g_{s}\geqslant 1/2 \\
\\
\textrm{for\quad }g_{s}<1/2%
\end{array}%
\right.   \label{av0}
\end{equation}%
The quantization condition is thus given by $a=-n$ for $\omega >0$ and $b=-n$
for $\omega <0$. Since $_{2}F_{1}\left( a,b,C,y\right) $ is invariant under
exchange of $a$ and $b$, one obtains a quantization condition independent of
the sign of $\omega $:

\begin{equation}
\mu +\nu +\frac{1-|\omega |}{2}=-n,\;n=0,1,2,\ldots  \label{el1}
\end{equation}%
Eq. (\ref{el1}) can also be written in the form

\begin{eqnarray}
&&\sqrt{\left( Mc^{2}\right) ^{2}-E^{2}}+\sqrt{\left(
Mc^{2}+g_{s}V_{0}\right) ^{2}-\left( E-g_{t}V_{0}\right) ^{2}}  \nonumber \\
&&  \label{el2} \\
&=&\frac{\hbar c}{2L}\left[ \sqrt{1+\left( 2g_{s}-1\right) \left( \frac{%
2LV_{0}}{\hbar c}\right) ^{2}}-\left( 2n+1\right) \right] ,\;n=0,1,2,\ldots
\nonumber
\end{eqnarray}%
Now we have an irrational algebraic equation to be solved numerically, but
there are still some questions that one ought to get answered. Does it
furnish proper solutions for the KG equation? Evidently $\varphi $ as in (%
\ref{phi1}) is a square-integrable function, and $\mu $ and $\nu $ must be
positive in order to furnish a wave function vanishing at $x=\pm \infty $.
Hence, the following supplementary conditions must be imposed:

\begin{equation}
\left\vert E\right\vert <|M|c^{2},\quad |E-g_{t}V_{0}|<|Mc^{2}+g_{s}V_{0}|
\label{ineq4b}
\end{equation}

\noindent as given by (\ref{ineq4}), and

\begin{equation}
g_{s}>1/2,\quad n=0,1,2,...<s  \label{sup1}
\end{equation}

\noindent where

\begin{equation}
s=\frac{1}{2}\left[ -1+\sqrt{1+\left( 2g_{s}-1\right) \left( \frac{2LV_{0}}{%
\hbar c}\right) ^{2}}\right]  \label{esse}
\end{equation}%
The first pair of supplementary conditions ensures that $\mu $ and $\nu $
are positive. The second pair is necessary to make positive the right-hand
side of (\ref{el2}). Note that this last pair of supplementary conditions
imposes an additional restriction on the product $LV_{0}$ beyond that one
which makes $\omega $ a real number as given by (\ref{av0}). As a matter of
fact, those conditions kill all the possibilities for bound states if the
scalar coupling does not exceeds the vector coupling. At any rate, the
possible solutions of (\ref{el2}) constitute a finite set of solutions.
According to the second line of \ (\ref{sup1}) and (\ref{esse}) one has

\begin{equation}
LV_{0}>\hbar c\,\sqrt{\frac{n\left( n+1\right) }{2g_{s}-1}}  \label{num}
\end{equation}%
This means that the number of allowed bound states increase with $LV_{0}$,
and there is at least one solution, no matter how small is $LV_{0}$. Now the
Gauss hypergeometric series $_{2}F_{1}\left( a,b,C,y\right) $ reduces to
nothing but a polynomial of degree $n$ in $y$ when $a$ or $b$ is equal to $%
-n $: Jacobi's polynomial of index $\alpha $ and $\beta $. Indeed, for $a=-n$ one has
\cite{abr}

\begin{eqnarray}
_{2}F_{1}\left( a,b,C,y\right) &=&_{2}F_{1}\left( -n,\alpha +1+\beta
+n,\alpha +1,y\right)  \nonumber \\
&&  \label{jaco1} \\
&=&\frac{n!}{\left( \alpha +1\right) _{n}}P_{n}^{\left( \alpha ,\beta
\right) }\left( \xi \right)  \nonumber
\end{eqnarray}

\noindent where

\begin{equation}
\alpha =2\nu ,\quad \beta =2\mu ,\quad \xi =1-2y=\tanh \frac{x}{2L}
\label{alfabeta}
\end{equation}

\noindent and $\left( \alpha \right) _{n}=\alpha \left( \alpha +1\right)
\left( \alpha +2\right) ...\left( \alpha +n-1\right) $ with $\left( \alpha
\right) _{0}=1$. Jacobi's polynomials $P_{n}^{\left( \alpha ,\beta \right) }\left( \xi \right) $ are
orthogonal with respect to the weighting function $w^{\left( \alpha ,\beta
\right) }(\xi )=\left( 1-\xi \right) ^{\alpha }\left( 1+\xi \right) ^{\beta
} $ on the interval $\left[ -1,+1\right] $, and can be standardized as

\begin{equation}
P_{n}^{\left( \alpha ,\beta \right) }\left( 1\right) =\frac{\left( \alpha
+1\right) _{n}}{n!}  \label{stand}
\end{equation}

\noindent so that

\begin{equation}
\int_{-1}^{+1}d\xi \,\,w^{\left( \alpha ,\beta \right) }(\xi )P_{n}^{\left(
\alpha ,\beta \right) }\left( \xi \right) P_{n^{\prime }}^{\left( \alpha
,\beta \right) }\left( \xi \right) =\delta _{nn^{\prime }}h_{n}^{\left(
\alpha ,\beta \right) }  \label{orto}
\end{equation}

\noindent where $h_{n}^{\left( \alpha ,\beta \right) }$ is given by \cite%
{abr}

\begin{equation}
h_{n}^{\left( \alpha ,\beta \right) }=\frac{2^{\alpha +\beta +1}}{2n+\alpha
+\beta +1}\frac{\Gamma \left( n+\alpha +1\right) \Gamma \left( n+\beta
+1\right) }{n!\,\Gamma \left( n+\alpha +\beta +1\right) }  \label{hn}
\end{equation}

\noindent Hence the KG wave function can be written as (see Ref. \cite{nie}):

\begin{equation}
\varphi _{n}\left( \xi \right) =N_{n}\left( 1-\xi \right) ^{\alpha /2}\left(
1+\xi \right) ^{\beta /2}P_{n}^{\left( \alpha ,\beta \right) }\left( \xi
\right)   \label{17}
\end{equation}

\noindent where $N_{n}$ is given by

\begin{equation}
N_{n}=\sqrt{\frac{\alpha \beta }{2^{\alpha +\beta }L\left( \alpha +\beta
\right) }\,\frac{\Gamma \left( \alpha +\beta +n+1\right) \Gamma \left(
n+1\right) }{\Gamma \left( \alpha +n+1\right) \Gamma \left( \beta
+n+1\right) }}  \label{19}
\end{equation}

\noindent in such a manner that%
\begin{equation}
\int_{-\infty }^{+\infty }dx\,\left\vert \varphi _{n}(x)\right\vert ^{2}=1
\label{n1}
\end{equation}

\noindent and

\begin{equation}
\int_{-\infty }^{+\infty }dx\,V_{t}(x)\left\vert \varphi _{n}(x)\right\vert
^{2}=V_{0}\,\frac{\beta }{\beta +\alpha }  \label{n2}
\end{equation}%
Nevertheless, using (\ref{normarho}) and (\ref{ineqE}) one can show that the
normalized KG wave function must be written as

\begin{equation}
\varphi _{n}^{\left( N\right) }(\xi )=N_{\pm }\varphi _{n}(\xi )  \label{f1}
\end{equation}

\noindent with

\begin{equation}
N_{\pm }=\sqrt{\,\pm \,mc^{2}\,\frac{\alpha +\beta }{\left( \alpha +\beta
\right) E-g_{t}V_{0}\beta }},\quad \quad \pm \quad \mathrm{for\,}\quad
E\gtrless g_{t}V_{0}\,\frac{\beta }{\beta +\alpha }  \label{f2}
\end{equation}%
It is worthwhile to note that invoking the property $P_{n}^{\left( \alpha
,\beta \right) }\left( -\xi \right) =(-1)^{n}P_{n}^{\left( \beta ,\alpha
\right) }\left( \xi \right) $, one can conclude that the KG wave functions
with definite parities (putting $\alpha =\beta $), associated with an even
potential, are obtained on the condition that the uniform background
satisfies (\ref{vezinho}), i.e. the effective potential is the symmetric
Rosen-Morse potential.

\section{Conclusions}

We have explored the influence of a scalar uniform background added to a
mixed vector-scalar smooth step potential. We have verified that the
background presents drastic effects on both scattering and bound-state
solutions. The background increases the threshold for the pair production
when the vector coupling exceeds the scalar coupling and $\mathcal{V}%
>-mc^{2}/g_{s}$. On the other hand, if $\mathcal{V}<-mc^{2}/g_{s}$ the
background decreases the threshold and pair production may occur for a
potential barrier arbitrarily small. When the scalar coupling exceeds the
vector coupling there appears the possibility of a finite set of bound-state
solutions. It is curious that the smooth step potential might hold bound
states in spite of the fact that the potential given by (\ref{4}) is
everywhere repulsive, so that one can not expect bound states in the
nonrelativistic limit. Of course, the scalar uniform background plays a
peremptory role for the actual occurrence of bound states but no
nonrelativistic limit can expected since the background has to be in a range
of values which do not acquiesce a nonrelativistic limit.

\bigskip

\bigskip

\noindent \textbf{Acknowledgments}

This work was supported in part through funds provided by CNPq.

\end{document}